\documentclass[journal]{IEEEtran}

\usepackage{amsfonts,amssymb,amsmath,amsthm}
\usepackage{algorithm}
\usepackage{graphicx}
\usepackage{xcolor}

\newcommand{\bH}{\mathbf{H}}
\newcommand{\z}{\mathbf{z}}
\newcommand{\y}{\mathbf{y}}
\newcommand{\m}{\mathbf{m}}
\newcommand{\w}{\mathbf{w}}

\newcommand{\I}{\mathbf{I}}
\newcommand{\C}{\mathbf{C}}
\newcommand{\bP}{\mathbf{P}}
\newcommand{\B}{\mathbf{B}}
\newcommand{\bB}{\mathbf{B}}
\newcommand{\M}{\mathbf{M}}

\renewcommand{\(}{\left(}
\renewcommand{\)}{\right)}
\newcommand{\al}{\boldsymbol{\alpha}}
\title{Probabilistic Simplex Component Analysis
by Importance Sampling}

\author{Nerya Granot, Tzvi Diskin, Nicolas Dobigeon and Ami Wiesel\thanks{The first two authors equally contributed to this letter. The research was partially supported by ISF grant number 2672/21, the ANR-3IA Artificial and Natural Intelligence Toulouse Institute (ANITI) under grant agreement ANITI ANR-19-PI3A-0004 and the ANR Active Molecular Imaging and Unmixing (ANR IMAGIN) Project under Grant ANR-21-CE29-0007.}} \date{\empty}

\begin{document}

\maketitle

\begin{abstract} 
In this paper we consider the problem of linear unmixing hidden random variables defined over  the simplex with additive Gaussian noise, also known as probabilistic simplex component analysis (PRISM). Previous solutions to tackle this challenging problem were based on geometrical approaches or computationally intensive variational methods. In contrast, we propose a conventional expectation maximization (EM) algorithm which embeds importance sampling. For this purpose, the proposal distribution is chosen as a simple surrogate distribution of the target posterior that is guaranteed to lie in the simplex. This distribution is based on the Gaussian linear minimum mean squared error (LMMSE) approximation which is accurate at high signal-to-noise ratio. Numerical experiments in different settings demonstrate the advantages of this adaptive surrogate over state-of-the-art methods. 
\end{abstract} 


\IEEEpeerreviewmaketitle

\section{Introduction}

This letter  considers the problem of linear unmixing hidden random variables lying on the simplex corrupted by an additive Gaussian noise. The problem, recently coined as Probabilistic Simplex Component Analysis (PRISM) \cite{prism}, is a variant of Non-negative Matrix Factorization (NMF) \cite{Positive_matrix_factorization,faces} that assumes an underlying Dirichlet prior distribution on the mixing coefficients. This leads to a well defined and identifiable parameter estimation problem under the maximum likelihood paradigm. The main challenge is then to design a numerical solution to the underlying optimization that involves high dimensional marginalization over the latent variables. In line with other approaches already proposed in the literature, we propose to solve this problem by resorting to a particular instance of the popular Expectation Maximization (EM) algorithm. In particular, the a E-step is approximated by a Monte Carlo integrator based on importance sampling with a carefully designed proposal distribution.

PRISM and more generally linear unmixing have a rich history in the signal processing literature. Linear unmixing can be interpreted as a variant of NMF which demonstrate its interest in various applicative contexts including topic modeling \cite{topic_modeling_3} and hyperspectral imaging \cite{Hyperspectral_Unmixing}. Some geometry-inspired approaches formulates this task as recovering the simplex with the minimal volume that covers all of the samples \cite{craig1994minimum,li2008minimum,minimum_volume_chan}. Others propose to identify the  ``purest'' observations (e.g., pixels) associated with the vertices of the simple \cite{fu2019nonnegative,n_finder}. Methods have been derived for the noise-free case, for additive Gaussian noise and for more challenging scenarios involving outliers \cite{wu2017stochastic}. There is also a family of Bayesian solutions to this problem \cite{DECA,dobigeon2009baysian,dobigeon}. More advanced models also allow random mixing matrices to be characterized by  different types of distributions \cite{eches2010bayesian,woodbridge2019unmixing}. 

Closest to our letter is the recent PRISM paper which adopted a maximum likelihood formalism and derived its properties \cite{prism}. PRISM suggested two numerical solutions. The first ISA method based on importance sampling \cite{wei1990monte} was shown to be highly accurate but non-scalable. The second VIA method relied on variational inference using surrogate Dirichlet distributions, performed well in terms of accuracy and scalability, but was suboptimal at high signal-to-noise ratios (SNRs). These two methods motivate the present letter and are the building blocks to our proposed approach that unifies their ideas.

The main contribution of this letter is a normalized importance sampling approach to PRISM. First, we revisit ISA and show that using a simple surrogate based on the prior distribution, the resulting so-called SISA performs well even in large problems. Second, following VIA, we develop LISA, an adaptive importance sampling method. LISA uses Dirichlet surrogates based on the closed-form Linear Minimum Mean Squared Error (LMMSE) estimates. In a low SNR regime, LISA is shown to behave as SISA which is near optimal. At high SNR, LISA mimics the LMMSE estimate and provides its samples around the estimate.  Both SISA and LISA embed sampling schemes and are therefore computationally intensive. However, contrary to previous methods, their samples are guaranteed to lie within the simplex and thus are never rejected and ensure scalability. Numerical experiments using synthetic simulations demonstrate the advantages of the proposed methods. Results show that SISA can serve as a promising initialization to VIA and that LISA provides the best performance (especially in high SNR where VIA is theoretically suboptimal).

\section{Dirichlet preliminaries}\label{dir_properties}
Throughout this paper, we will focus on the simplex and its related Dirichlet distribution. Therefore, we begin with their definitions and basic properties. The $k$-dimensional simplex is defined as
\begin{align}
    \mathbb{S}_k = \left\{ \mathbf{z} \in R^k : z_i\geq 0,\;\mathbf{1}^T{\mathbf{z}}=1\right\},
\end{align}
where $\mathbf{1}$ is a length-$k$ vector of ones. A popular multivariate distribution over this simplex is the Dirichlet distribution whose probability density function (pdf) writes
\begin{align}
    {\rm{Dir}}(\mathbf{z};{\boldsymbol{\alpha}})\propto \prod_{n=1}^kz_n^{\alpha_n-1}, \quad {\mathbf{z}}\in \mathbb{S}_k,
\end{align}
where ${\boldsymbol{\alpha}}>0$ is the concentration parameter (the inequalities should be understood as a term-wise comparison). Its mean and covariance are given by
\begin{align}
\label{diri_mean_var}
    &\mathbf{m} = \frac{{\boldsymbol{\alpha}}}{\mathbf{1}^T{\boldsymbol{\alpha}}}\in \mathbb{S}_k,
    &\mathbf{C} = \frac{{\rm{diag}}(\mathbf{m})-\mathbf{mm}^T}{\mathbf{1}^T{\boldsymbol{\alpha}}+1},
\end{align}
and satisfy
\begin{align}\label{projection}
    \m = \mathbf{P}\m - \frac{1}{k}{\mathbf{1}}, \quad
     \mathbf{C}=\mathbf{PC}, \quad \mathbf{P}=\mathbf{I}-\frac{\mathbf{11}^T}{\mathbf{1}^T\mathbf{1}}.
\end{align}
Because of the linear dependence between the vector components, the covariance $\mathbf{C}$ is singular. 
\section{Problem formulation}
We consider linear mixing with random hidden variables, also known as PRISM \cite{prism}
\begin{equation}\label{unmixing_model}
    \mathbf{y_i}=\mathbf{Hz}_i+\mathbf{w}_i\quad i=1,\cdots,N
\end{equation}
where $\bH$ is a deterministic unknown matrix of size $d\times k$, $\z_i\sim p(\z)={\rm{Dir}}(\z;\al)$ are independent and identically distributed (i.i.d.) hidden random vectors from a Dirichlet distribution with a known deterministic parameter $\al$, and $\w_i$ are i.i.d. noise vectors ${\mathcal{N}}({\mathbf{0}},\sigma^2\I)$ with a known variance $\sigma^2$. We assume that $\z_i$ and $\w_i$ are independent. The goal is then to estimate $\bH$ given an observed set of measurements $\{\y_i\}_{i=1}^N$.\\
\noindent \textbf{EM algorithm --} A standard approach consists in maximizing the log-likelihood with respect to (w.r.t.) the unknown parameter \cite{kay}
\begin{align}\label{ML1}
    \hat \bH &= \arg\max_{\bH} \frac{1}{N}\sum_{i=1}^N\log p_\bH(\y_i).
\end{align}
The distribution of $\y$ is defined through the hidden variable $\z$ and requires marginalization 
\begin{align}
    p(\y)=\int p(\y|\z)p(\z)d\z.
\end{align}

Computing this high dimensional integral or its gradient is often impossible. A popular alternative is the EM algorithm that iteratively maximizes a lower bound \cite{em_by_dempster}. Each iteration of the overall algorithm consists of two steps. Given a current estimate ${\bH}'$ of the parameter, the first E-step computes a conditional expectation of the complete log-likelihood
\begin{equation}\label{eq:E_step}
({\rm{E}}) \quad Q(\bH;\bH') = \sum_{i=1}^N 
 {\rm{E}}[\log{p(\y_i|\z_i)}+\log{p(\z_i)}|\y_i;{\bH}'],
\end{equation}
where $\rm{E}[\cdot|\y;\bH']$ denotes the conditional expectation given $\y$ and $\bH'$. In the context of PRISM, the quantity \eqref{eq:E_step} can be explicitly derived as
\begin{align}
&\quad Q(\bH;\bH') =\sum_{i=1}^N\frac{\rm{E}[\|\bH\z_i-\y_i\|^2|\y_i;\bH']}{-2\sigma^2}\nonumber\\
&=\sum_{i=1}^N\frac{{\rm{ Tr}}(\bH^T\bH \rm{E}[\z_i\z_i^T|\y_i;\bH'])-2\y_i^T\bH E[\z_i|\y_i;\bH']}{-2\sigma^2}.
\end{align}
The second M-step searches for the parameter that maximizes this quantity
\begin{equation}
(\rm{M}) \quad \bH \leftarrow \max_\bH Q(\bH;\bH').
\end{equation}
It is easy to show that the EM algorithm can be cast as a minimization-majorization strategy \cite{sun2016majorization} and, under regular technical conditions, it is shown to converge to a stationary point of the log-likelihood. Combining the (E) and (M) steps, the EM iteration boils down to the updating rule
\begin{equation}
 \bH \leftarrow \sum_{i=1}^N\y_i{\rm{E}}^T[\z_i|\y_i;\bH']\left(\sum_{i=1}^N{\rm{E}}[\z_i\z_i^T|\y_i;\bH']\right)^{-1}.
\end{equation}
The main challenge with this strategy lies in the computation of $\rm{E}[\z_i|\y_i;\bH']$ and $\rm{E}[\z_i\z_i^T|\y_i;\bH']$ efficiently for each sample at each iteration. One solution consists in resorting to a Monte Carlo approximation, resulting in a so-called Monte Carlo EM algorithm \cite{wei1990monte}.\\
\noindent \textbf{MCEM algorithm --} A classical technique for Monte Carlo approximations is referred to as importance sampling (IS) which relies on a surrogate (or proposal) distribution $q(\cdot)$ it is easier to sample from. 
In the context of PRISM, the task can be formulated as computing quantities of the form
\begin{align}\label{eq:integral}
    {\rm{E}}[d(\z,\y)|\y] &=\int d(\z,\y)p(\z|\y)d\z \\
        &=\frac{1}{C(\y)} \int d(\z,\y)p(\y|\z)p(\z)d\z \nonumber
\end{align}
where $C(\y)= \int p(\y|\z)p(\z)d\z$  and $d(\cdot,\y)$ specifies the quantity of interest. Under generic assumptions about the proposal $q(\cdot)$, the expectation in \eqref{eq:integral} can be rewritten as
\begin{align}
    {\rm{E}}[d(\z,\y)|\y] &=\frac{1}{C} \int d(\z,\y)p(\y|\z)p(\z)\frac{q(\z|\y)}{q(\z|\y)}d\z.\nonumber    
    \end{align}
Then, for a given set of $M$  i.i.d. samples  $\z_m$ drawn from $q(\z|\y)$, IS proceeds with a Monte Carlo approximation
\begin{equation}\label{IS}
{\rm{E}}[d(\z,\y)|\y] \approx \frac{1}{\tilde{C}} \sum_{m=1}^M \tilde{w}_m d(\z_m,\y)
    \end{equation}
with $\tilde{w}_m = \frac{p(\y|\z_m)p(\z_m)}{q(\z_m|\y)} $ and $\tilde{C}=\sum_{m=1}^M \tilde{w}_m$. 
 The quality of the approximation \eqref{IS} is governed by the similarity between the target distribution and its surrogate. The goal is therefore to choose a surrogate distribution $q(\z|\y)$ which is a good approximation to $p(\z|\y)$ and easy to sample from. This point is the core of the next section.

\section{Surrogate posterior distributions}
This section discusses several choices of accurate yet cheap surrogates $q(\z|\y)$ approximating $p(\z|\y)$ for $\z\in \mathbb{S}_r$. 
\subsection{Dirichlet prior}\label{sec_sisa}
The simplest surrogate distribution, denoted by Simple ISA (SISA), ignores $y$ and approximates the posterior by the prior 
\begin{align}
  q(\z|\w)=p(\z)={\rm{Dir}}(\z;\al).
\end{align}
This approach should be optimal for low SNR where $p(\z|\y)\approx p(\z)$. Otherwise, it seems wasteful as it ignores the information brought by $\y$. SISA can also be derived as a Sample Average Approximation or naive Monte Carlo averaging \cite{wu2017stochastic}.

\subsection{Gaussian posterior}
The target posterior distribution $p(\z|\y)$ is a multivariate Gaussian distribution truncated on the simplex. One solution to generate samples from this distribution consists in resorting to rejection sampling. Such a strategy is shown to performed poorly for large values of $k$ due to a high rejection rate. In particular, the authors of \cite{prism} stated, in settings identical to those in our experiments, this method generated almost no samples. One alternative would rely on more advanced Monte Carlo techniques, e.g., Markov Chain Monte Carlo (MCMC) algorithms \cite{Altmann_IEEE_SSP_2014}. However, such strategies are generally computationally demanding and can be hardly embedded into the iterative scheme of EM.

One alternative is the conditional Gaussian distribution, also known as Linear Minimum Mean Squared Error (LMMSE) estimation, denoted by ${\mathcal{N}}(\z;\overline{\m}(\y),\overline \C)$ with
\begin{align}\label{gauss_moments}
    &\overline \m(\y) = \m + \C\bH^T(\bH\C\bH^T+\sigma^2\I)^{-1}(\y-\bH\m)\nonumber\\
    &\overline \C = \C - \C\bH^T(\bH\C\bH^T+\sigma^2\I)^{-1}\bH\C,
\end{align}
where $\m$ and $\C$ are the prior Dirichlet moments. This approximation is near optimal in high SNR regimes where the $\overline \m(\y)\approx \z$ is accurate and it makes sense to sample around it. 
Unfortunately, with even small noise, samples from this distribution do not necessarily lie in the simplex and this approach leads to a high rejection rate. 

\subsection{Dirichlet posterior}\label{sec_lisa}
A more promising approximation, denoted by LISA, relies on the Dirichlet distribution but adjusts it according to LMMSE estimate.
We define 
\begin{align}
    \overline \z\sim q(\overline \z|\y)={\rm{Dir}}(\overline \z;\overline\al(\y)),
\end{align}
which is guaranteed to lie in the simplex and choose $\overline{\al}(\y)$ to fit the moments in (\ref{gauss_moments}), i.e., 
\begin{align} 
    &{\rm{E}}[\overline \z]=\widetilde \m(\y) \label{eq:LISA_constraints_1} \\
    &{\rm{Tr}}[{\rm{cov}}[\overline \z]]={\rm{Tr}}[\overline \C] \label{eq:LISA_constraints_2},
\end{align}
where $\widetilde \m(\y)$ is the (approximate) projection of $\overline \m(\y)$ onto $\mathbb{S}_k$. Indeed, the Dirichlet mean is always within the simplex and since it has $k-1$ degrees of freedom, imposing \eqref{eq:LISA_constraints_1} and \eqref{eq:LISA_constraints_2} boils down to fit $k$ parameters to $k$ moments constraints . The first moment constraint
\begin{align}
    {\rm{E}}[\overline \z]=\frac{\overline{\boldsymbol{\alpha}}}{{\mathbf{1}}^T\overline{\boldsymbol{\alpha}}}=\widetilde \m(\y),
\end{align}
leads to $\overline\al = \mu \widetilde \m(\y)$ with some $\mu >0$. The scaling factor $\mu$ controls the variance. It is adjusted to enforce the covariance
\begin{align*}
    {\rm{Tr}}[{\rm{cov}}[\overline \z]]={\rm{Tr}}\left[\frac{{\rm{diag}}(\widetilde \m(\y))-\widetilde \m(\y)\widetilde \m^T(\y)}{\mu+1}\right]={\rm{Tr}}(\overline \C),
\end{align*}
which yields
\begin{align}
    \mu = \frac{1-\|\overline \m(\y)\|^2}{\rm{Tr}(\overline \C)}-1.
\end{align}
To summarize, the LISA proposal distribution is defined as
\begin{align}
    q(\overline \z|\y)={\rm{Dir}}\left(\overline \z; \left(\frac{1-\|\widetilde \m(\y)\|^2}{{\rm{Tr}}(\overline \C)}-1\right)\widetilde \m(\y)\right).
\end{align}
Capitalizing on the properties stated in Section \ref{dir_properties}, one can easily characterize the asymptotic behavior of this proposal wrt to the noise level. In low SNR, LISA depends only on the prior and we get $\widetilde \m(\y)\approx \m$ and $\overline{\C} \approx \C$. After some algebraic manipulations this yields
\begin{align}
    & q(\overline \z|\y) \stackrel{{\text{low\;SNR}}}{\rightarrow} {\rm{Dir}}\left(\overline \z;\al\right),
\end{align}
so that LISA converges to SISA in low SNR. Conversely, in high SNR, LISA does not depend on the prior $\al$. The moments reduce to (see the Appendix for proof):
\begin{align}
    &\widetilde \m(\y)\stackrel{{\text{high\;SNR}}}{\rightarrow} (\bH\bP)^\dag \y + \mathbf{v}_\bH\nonumber\\
&\overline{\C}\stackrel{{\text{high\;SNR}}}{\rightarrow} \sigma^2(\bP\bH^T\bH\bP)^{\dag}, \label{high SNR}
\end{align}
where $\mathbf{v}_\bH = \frac{1}{k}\left(\I- \left(\bH\bP\right)^\dagger\bH\right)\mathbf{1}$ and we get
\begin{align}
    & q(\overline \z|\y) \stackrel{{\text{high\;SNR}}}{\rightarrow} {\rm{Dir}}\left(\overline \z;\frac{c}{\sigma^2}[(\bH\bP)^\dag \y+\mathbf{v}_\bH]\right),
\end{align}
where $c>0$ is a constant. As expected, this yield samples which are concentrated around the LMMSE estimate with a small variance.

\begin{figure*}
\includegraphics[width=0.32\textwidth]{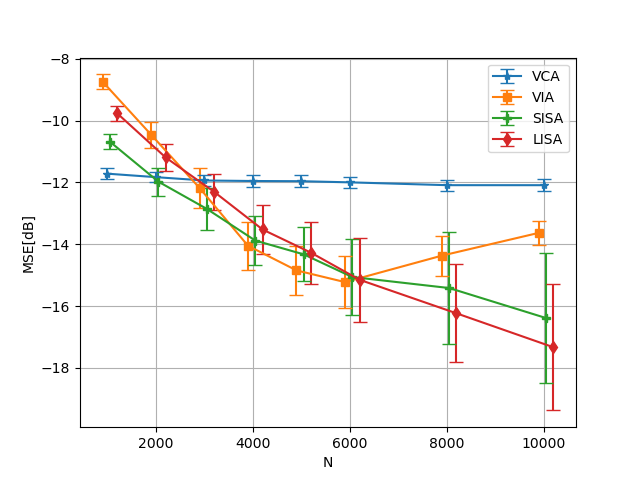}\includegraphics[width=0.32\textwidth]{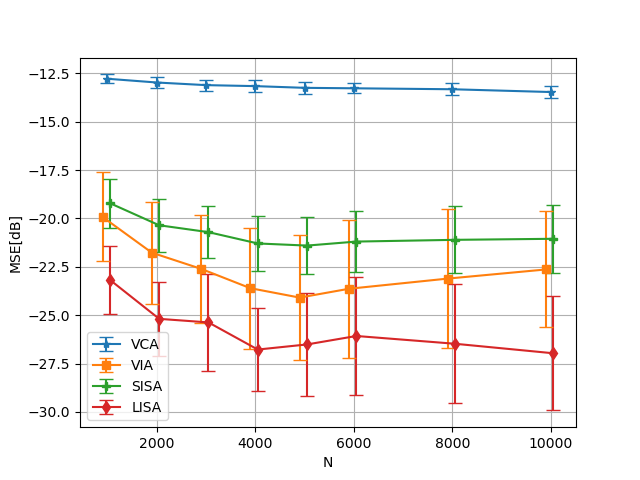}
\includegraphics[width=0.32\textwidth]{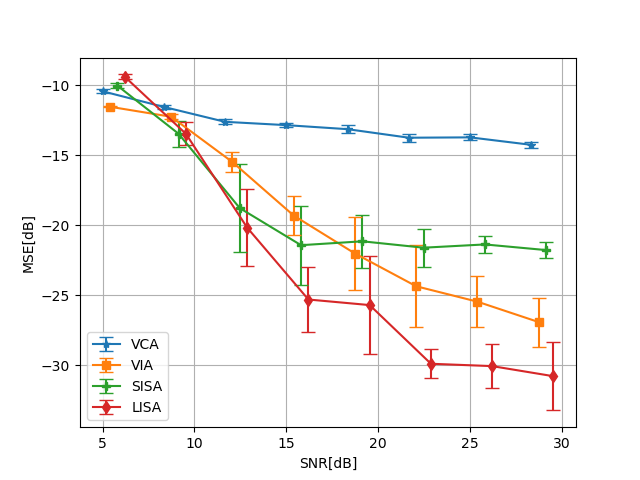}

	\caption{MSE as a function of the number of samples in $\mathrm{SNR}=10{\rm dB}$ \textbf{(left)}, the number of samples in $\mathrm{SNR}=20{\rm dB}$ \textbf{(middle)} and the SNR with $N=5000$ \textbf{(right)}. The SNR is given by ${\rm Tr}\(\bH\C\bH^T\)/\sigma^2$.}
   \label{fig:snrs}
\end{figure*}
\section{Numerical experiments}
This section compares the performance of the different algorithms using numerical experiments. The simulations are reproduction of the synthetic experiments in \cite{prism} with the exact settings. The data were generated synthetically based on the linear unmixing model in (\ref{unmixing_model}). The matrix $\bH$ of dimensions $d=50$ and $k=20$ was generated once per experiment with i.i.d. elements uniformly distributed in $[0,1]$. 
Performance was measured by mean squared error over the best permutation:
\begin{align}
    {\rm{MSE}}=\min_{\pi\in\Pi}\sum_i\|\bH_i-\hat \bH_{\pi_i}\|^2,
\end{align}
where $\bH_i$ is the $i$the column of $\bH$ and $\Pi$ is the set of all indices permutations. Four competing algorithms are compared
\begin{itemize}
    \item VCA: A simple and fast baseline \cite{vca}.
    \item SISA: An EM method initialized by VCA and using standard importance sampling as detailed in Sec. \ref{sec_sisa}. The EM has $100$ iterations and SISA is based on $M=500$ samples. 
    \item LISA: A similar EM method where the last 50 iterations use an LMMSE surrogate as detailed in Sec. \ref{sec_lisa}.
    \item VIA: A variational approach due to \cite{prism}. Following \cite{VIA_next}, we implemented VIA using Torch with a line search for the learning rate. In order to achieve good accuracy, we initialized VIA with SISA.
\end{itemize}
The first experiment considered performance as a function of the number of samples in low SNR. The results are provided in Fig. \ref{fig:snrs}. As expected from the theory, SISA was near optimal in low SNR. LISA behaved similarly and outperformed it when the number of samples is large.

The second experiment repeated the experiment in higher SNR. The results are provided in Fig. \ref{fig:snrs}. This setting is more challenging for SISA which is outperformed by VIA. LISA was significantly better than the rest of the algorithms throughout this graph. 

The third experiment in Fig. \ref{fig:snrs} examined the performance for a fixed number of samples $5000$ as a function of the SNR. Here too the advantages of LISA are apparent. It is only to see the expected degradation in performance of VIA in high SNR.
In terms of computational complexity, VCA is the fastest algorithm. SISA and LISA are significantly higher because of the sampling. LISA is slightly more expensive than SISA because of its data dependent concentration parameters. Finally, VIA is more tricky. The original implementation in \cite{prism} is quite complicated, but the Torch implementation of \cite{VIA_next} is very fast. However, in order to get the performance detailed above, we had to initialize VIA with SISA and this slowed it down considerably.

\appendix[Derivation of LISA in high SNR (\ref{high SNR})]

First, we show that if $\bH$ is full rank and $d\geq k$, then  
\begin{equation}\label{limit}
\M=\lim_{\sigma^2\rightarrow0}\C\bH^T\left(\bH\C\bH^T+\sigma^2\I\right)^{-1}=\left(\bH\mathbf{P}\right)^\dagger.
\end{equation}
$\C$ is positive semi-definite and there exists a matrix $\bB$ such that $\bB\bB=\C$: 
\begin{align}
\M&=\lim_{\sigma^2\rightarrow0}\B\B\bH^T\left(\bH\B\B\bH^T+\sigma^2\I\right)^{-1}=\B\left(\bH\B\right)^{\dagger}.
\end{align}
The null space of $\bB$ is the same as $\C$ and $\bP$, and therefore $\bB^\dagger \bB = \bP$. It remains to prove that $\M'^\dagger=\bH\B\B^\dagger$ is the pseudo-inverse of $\M=\B\left(\bH\B\right)^\dagger$ using the four Moore-Penrose conditions \cite{golub2013matrix}:


(I) 
 Because $\bH$ is full rank and $n\geq k$, $\bH^\dagger\bH=\I_{k\times k}$ and:
\begin{align}\label{33}
    \M\M'^\dagger &=  \B\left(\bH\B\right)^\dagger\bH\B\B^\dagger = \bH^\dagger\bH\B\left(\bH\B\right)^\dagger\bH\B\B^\dagger \nonumber \\
    &=\bH^\dagger\bH\B\B^\dagger=\B\B^\dagger,
\end{align}
which is clearly a symmetric orthogonal projection. (II) Similarly,
\begin{align}\label{34}
    \M'^\dagger\M &=\bH\B\B^\dagger \B\left(\bH\B\right)^{\dagger}=\bH\B\left(\bH\B\right)^\dagger,
\end{align}
which is again symmetric. (III) Using (\ref{33}), 
\begin{align}
    \M\M'^\dagger\M &=\B\B^\dagger\B\left(\bH\B\right)^{\dagger} = \B\left(\bH\B\right)^{\dagger}=\M. 
\end{align}
(IV) Finally, using (\ref{34}), we have
\begin{align}
    \M'^\dagger\M \M'^\dagger &=\bH\B\left(\bH\B\right)^\dagger \bH\B\B^\dagger =  \bH\B\B^\dagger=\M'^\dagger.
\end{align}

Next, we show the mean terms yield $\mathbf{v}_\bH$ which is independent of the prior ${\boldsymbol{\alpha}}$. 
Plugging (\ref{limit}) into (\ref{gauss_moments}) gives:
\begin{align}\label{36}
    \overline \m(\y) &= \m + \left(\bH\bP\right)^\dagger\left(\y-\bH\m\right) \nonumber \\
    &=\left(\bH\bP\right)^\dagger\y + \m - \left(\bH\bP\right)^\dagger\bH\m.
\end{align}
Now using the fact that $\m$ satisfies (\ref{projection}), we get:
\begin{align}\label{37}
    \left(\bH\bP\right)^\dagger\bH\m &= \left(\bH\bP\right)^\dagger\bH\left(\bP\m + \frac{\mathbf{1}}{k}\right) \nonumber \\
    &=\left(\bH\bP\right)^\dagger\bH\bP\m +\frac{1}{k}\left(\bH\bP\right)^\dagger\bH \mathbf{1}.
\end{align}
We note that because $\bH$ is full rank, the null space of $\bH\bP$ is the same as of $\bP$ and thus $ \left(\bH\bP\right)^\dagger\bH\bP = \bP $.
Therefore,
\begin{align}\label{38}
    \left(\bH\bP\right)^\dagger\bH\bP\m = \bP\m= \m -\frac{\mathbf{1}}{k}.
\end{align}
Plugging (\ref{38}) into (\ref{37}) and then to (\ref{36}) gives:
\begin{equation}
    \overline \m(\y) = \left(\bH\bP\right)^\dagger\y +\frac{1}{k}\left(\I- \left(\bH\bP\right)^\dagger\bH\right)\mathbf{1}.
\end{equation}

Finally, the covariance of the error is given by:
\begin{align}
    \overline{\C} &= {\rm cov}(\hat \z-\z) = {\rm cov}\left(\left(\bH\bP\right)^\dagger\left(\y-\bH v_\bH)\right)-\z\right) \nonumber \\
    & = {\rm cov}\left(\left(\left(\bH\bP\right)^\dagger\bH-\I\right)\z\right) + {\rm cov}\left(\left(\bH\bP\right)^\dagger {\mathbf{w}}\right) \nonumber\\
    & = \left(\left(\bH\bP\right)^\dagger\bH-\I\right)\C\left(\left(\bH\bP\right)^\dagger\bH-\I\right)^T \nonumber \\
    &+ \sigma^2\left(\bH\bP\right)^\dagger{\left(\bH\bP\right)^\dagger}^T.
\end{align}

Now we note that
\begin{align}
    &\left(\left(\bH\bP\right)^\dagger\bH-\I\right)\C =\left(\left(\bH\bP\right)^\dagger\bH-\I\right)\bP\C \nonumber \\
    &=\left(\left(\bH\bP\right)^\dagger\bH\bP-\bP\right)\C 
    =\left(\bP-\bP\right)\C = \mathbf{0},
\end{align}
and thus:
\begin{equation}
    \overline{\C} = \sigma^2\left(\bH\bP\right)^\dagger{\left(\bH\bP\right)^\dagger}^T =\sigma^2\left(\bP\bH^T\bH\bP\right)^\dagger.
\end{equation}

\bibliographystyle{IEEEtran}
\bibliography{sample}
\end{document}